\documentstyle[aps,draft]{revtex}
\def\lvert{\hspace{0.30ex}|\hspace{0.30ex}}
\def\ket{\hspace{0.30ex} \rangle}
\def\bra{\langle\hspace{0.30ex}}
\newcommand{\btau}{\mbox{\boldmath $\tau$}}

\begin{document}
%\jl{1}

%\draft
\title{Relationship between long time scales and \\ the static
free-energy in the Hopfield model}
\author{G.Biroli and R.Monasson
\footnote{Laboratoire de Physique Th\'eorique de l'ENS; preprint
number 98/04}}
\address{CNRS-Laboratoire de Physique Th{\'e}orique de l'ENS,\\
24 rue Lhomond, 75231 Paris cedex 05, France}

\maketitle
\begin{abstract}
The Glauber dynamics of the Hopfield model at low storage level is
considered. We analytically derive the spectrum of relaxation times
for large system sizes. The longest time scales are gathered in
families, each family being in one to one correspondence with a
stationary (not necessarily stable) point of the static mean-field
free-energy. Inside a family, the time scales are given by the
reciprocals (of the absolute values) of the eigenvalues of the 
free-energy Hessian matrix. 
\end{abstract}

%\narrowtext

Over the past few years, there has been a renewal of interest for the
long time dynamics of spin-glass models, which may already exhibit non
trivial features, e.g. violation of fluctuation-dissipation relations,
aging phenomena, ... at the mean-field level \cite{revue}.  In this
context, a natural quantity of interest is the spectrum of relaxation
times occuring in the Glauber dynamics of disordered Ising spin models
\cite{glauber}.  Unfortunately, numerical investigations have been so
far limited to very small sizes \cite{melin}. Moreover, to the best of
our knowledge, no analytical studies of the relaxation times spectra
have been performed yet due to the technical difficulties arising in
the diagonalization of the Glauber matrix.

In this letter, we focus on the Glauber dynamics of the Hopfield model
at low storage \cite{hopfield}. This system is simple enough to be
analytically solved along the lines of \cite{dinamicaising}.  Our main
result is that the longest time scales are gathered in families, each
family being in one to one correspondence with a stationary (not
necessarily stable) point of the static mean-field free-energy. Inside
a family, the time scales are given by the reciprocals of the absolute
values of the eigenvalues of the free-energy Hessian matrix.  As a
consequence, our study strenghtens the close relationship between long
times dynamics and static properties in disordered mean-field models
\cite{revue}.

We consider a Hopfield model including $N$ Ising spins $S_i$,
$i=1,\ldots , N$ and $p$ quenched patterns $\xi _i ^{\mu}$, $\mu = 1
,\ldots , p$. In addition to its intrinsic interest as a paradigm for
attractor neural network \cite{amit}, the Hopfield model may be seen
as a spin glass system smoothly interpolating between the Mattis model
\cite{Mattis} (when $p=1$) and the much more involved
Sherrington-Kirpatrick model (for infinite $p$) \cite{sk}. Its Hamiltonian
depends on the spin configuration ${\bf S} = ( S_1 ,\ldots , S_N)$
through the set of $p$ overlaps ${\displaystyle m^{\mu}({\bf S})=\frac
1N \sum_{i=1}^{N}S_{i} \xi ^\mu _i}$ and reads
\begin{equation}
\label{hami}
H({\bf S})= -\frac{N}{2}\sum_{\mu =1}^{p}m^{\mu
}({\bf S})^{2} 
\end{equation} 
The thermodynamics of the above Hamiltonian has been studied in
\cite{amit}. At low storage level, that is when $p$ keeps finite as
the size $N$ is sent to infinity, the free-energy density is
particularly simple, 
\begin{equation}
\label{freeenergy}
f(\beta )= - \frac{1}{2} \sum_{\mu =1}^{p}(m^{\mu })^{2} +
\frac{1}{\beta } \ll 
\ln \left[ 2 \cosh \left( \beta  \sum_{\mu =1}^{p} m^{\mu }\xi ^{\mu}
\right) \right] \gg
\quad ,
\end{equation}
where $\beta$ is the inverse temperature and ${\displaystyle \ll F(\{
\xi ^\mu \} )\gg \equiv \frac 1N \sum _{i=1}^ N F(\{ \xi _i ^\mu \})}$
denotes the site average. In the above expression (\ref{freeenergy}),
the overlaps are the solutions of the saddle-point equations
\begin{equation}
\label{mfeq} 
m^{\mu }=\ll \xi
^{\mu }\tanh \left(\beta \sum_{\nu =1}^{p}  m^{\nu }\xi ^{\nu }
\right) \gg \quad ,
\end{equation} 
with the smallest free-energy. At high temperature, the system lies in
the paramagnetic phase $m ^ \mu =0$, $\forall \mu $. Below the
critical temperature $1/\beta _c =1$, there appear many other
solutions with non zero overlaps. The low temperature phase is
characterized by a condensation along some particular patterns, while
spurious mixtures give rise to metastable states.

These equilibrium states may be seen as the attractors of a retrieval
dynamics \cite{glauber} starting from a given initial configuration
(which is often but not necessarily chosen as a corrugated version of
the quenched patterns). One possible evolution scheme is provided by
the usual Glauber dynamics~: given a configuration ${\bf S}$, the spin
at site $i$ is flipped with probability rate $w({\bf S} ,i
)=\frac{1}{2}(1-S_{i}\tanh [\beta h_{i}({\bf S})] )$ and left
unchanged with rate $1-w$.  The instantaneous magnetic field
$h_{i}$ reads from (\ref{hami}), 
\begin{equation}
\label{champ}
h_i({\bf S})=\sum_{j (\ne i) } \left( \frac 1N \sum_{ \mu = 1}^p 
\xi _i ^\mu \xi_j^\mu \right) S_j  =\sum_{\mu = 1 }^p \xi_i^\mu m^\mu ({\bf
S})-\frac{p}{N}\; S_i
\end{equation}
The probability distribution $P({\bf S}, t)$ satisfies the master equation
\begin{equation}
\label{masterequation}
\frac{d}{dt} P({\bf S},t)=\sum_{i=1}^{N} \big( 
w({\bf S}^i, i)\; P({\bf S}^i,t)\; -\; w({\bf S}, i) \;P({\bf S},t)
\big) \quad ,
\end{equation}
where the spin configuration ${\bf S}^i$ is obtained from ${\bf S}$ by
flipping spin $i$. Due to its linear structure,
eqn. (\ref{masterequation}) may be recast in a more suitable operator
formalism. To each configuration ${\bf S}$ is associated a vector
$\lvert {\bf S} \ket$, \ which form the basis of a $2^N$-dimensional
linear space ${\cal V}$. The probability distribution then becomes a
vector ${\displaystyle \lvert P(t)\ket = \sum_{\bf S} P({\bf
S},t)\lvert {\bf S} \ket}$. We now define the spin flip and
magnetization operators, respectively denoted by $\sigma _i ^x$ and
$\sigma _i ^z $ through $\sigma _i ^x \lvert {\bf S} \ket = \lvert
{\bf S} ^i \ket$ and $ \sigma _i ^z \lvert {\bf S} \ket = S_i \lvert
{\bf S} \ket$.  It is easy to check that the vectorial operator $\frac
12 {\mbox{\boldmath $\sigma$}} = \frac 12 (\sigma ^x , \sigma ^y
\equiv i \sigma ^x \sigma ^z , \sigma ^z )$ satisfies the usual
commutation relations for a quantum spin $1/2$.  The master equation
(\ref{masterequation}) now acquires the following compact form
\begin{equation}
\frac{d}{dt} \lvert
P(t)\ket = W \lvert P (t)\ket \qquad , \qquad 
W =\frac{1}{2} \sum_{i=1}^N \big(\sigma _{i}^{x}-1 \big)
\big(1-\sigma _{i}^{z}\tanh(\beta h_{i}) \big) \quad ,
\end{equation}
where the local field operator $h_i$ is obtained by
replacing in (\ref{champ}) the spins $S_i$ with their corresponding 
magnetization operators $\sigma_i ^z$. 
Due to the detailed balance conditions, the operator 
$W^s = \exp ( \beta H /2 ) W \exp ( -\beta H /2 )$ is Hermitian and reads
\begin{equation}
\label{mastersym}
W^s =-\frac N2 + \frac 12 \sum_{i=1}^{N} \bigg[ \sigma _{i}^{x} \; ( \cosh
(\beta h_i) )^{-1} +\sigma _i ^z \tanh(\beta h_i ) \bigg] \qquad .
\end{equation}
The spectrum of $W^s$, which is the same as $W$'s one, we shall now
analyze. From a rigorous point of view, all eigenvalues are strictly
negative except the zero mode corresponding to the equilibrium Gibbs
measure.  First, as far as long time dynamics is concerned, all large
(in absolute value) eigenvalues, that are associated to fast decaying
modes, may be discard.  Secondly, for very large system sizes
($N\gg 1$), indeed the limit of physical interest, the zero
eigenvalue of $W^s$ is expected to become almost degenerate when $\beta >
\beta _c$. Dynamical processes taking place over exponentially
large times, or more generally times diverging with $N$ will not be
considered. We therefore focus only on long but finite time scales.

To do so, we shall perform an expansion of $W_s$ in powers of $1/\sqrt
N$ and keep only the non vanishing terms in the thermodynamical
limit. This expansion is made possible by the mean-field nature of the
Hopfield Hamiltonian (\ref{hami}). For a given configuration $\bf S$,
the dominant contributions $H_i \equiv \displaystyle {\sum _{\mu} \xi
_i ^\mu m ^\mu ({\bf S }) }$ to the local fields $h_i$ (\ref{champ})
take only $2^{p}$ different values when scanning all sites $i$. Let us
partition the $N$ sites in $2^p$ blocks $G_{\btau}$ where ${\btau}$ 
is a $p$ binary component vector; site $i$ belongs to the only
group $G _{\btau}$ such that $H_i$ equals $H _{\btau} =
\displaystyle {\sum _{\mu} \tau ^\mu m ^\mu ({\bf S }) }$ \cite{hem}. We
restrict to quenched patterns that differ from each other on a 
macroscopic number of sites, so each block $G _{\btau}$ 
contains a number $N c_{\btau}$ of sites of the order of $N$; note
that if the patterns are randomly drawn from an unbiased distribution
$c _{\btau}=2^{-p}$ in the limit of large sizes.

After some
simple algebra, the evolution operator may be rewritten as $W^s=
\displaystyle{\sum _{\btau} W^s _{\btau}}$ with
\begin{equation}
\label{Wdev}
W^s _{\btau}= -\frac N2 c_{\btau} + 
J ^x _{\btau} \; (\cosh(\beta H _{\btau}
))^{-1} +J ^{z} _{\btau}\; \tanh(\beta
H _{\btau} )-\frac{\beta p}{2} c _{\btau} (1 - \tanh (\beta H_{\btau} 
)^2 ) + O \left( \frac{ J^{y} _{\btau}}{N} \right) 
\end{equation}
where the angular momenta and the field operators respectively read
\begin{equation}
\label{cinet}
{\bf J}_{\btau } = \frac{1}{2} \sum_{i\in G_{\btau} } 
{\mbox{\boldmath $\sigma$}}
_{i} \qquad \hbox{\rm and}  \qquad 
H _{\btau} = \frac 2N \sum _{\mbox{\boldmath $\eta$}}
\sum _{\mu =1 }^p \tau ^\mu \eta ^\mu \; J _{\mbox{\boldmath $\eta$}} ^z
\quad .
\end{equation}
The last two terms on the r.h.s of (\ref{Wdev}) stem from the
contributions $p S_i /N$ to the local fields (\ref{champ}). The
angular momenta ${\bf J}_{\btau }$ and ${\bf J} _{{\btau }'}$
commute for any two different groups ${\btau} \ne
{\btau }' $ and so does any ${\bf J} _{\btau }$ with $W^s$. Thus,
the angular momenta may be used as quantum numbers labelling the
different invariant subspaces.

It can be checked that the subspace ${\cal V}^s$ corresponding to the
maximum value of each angular momentum $J_{\btau}= \frac 12 N
c_{\btau}$ contains the totally symmetric vectors $\lvert \{ M_{\bf
\tau} \} \ket$ having magnetizations $M_{\btau} = -N c_{\btau}, - N
c_{\btau} +2 , \ldots , N c_{\btau}$ in blocks $G_{\btau}$.  Indeed,
the vector $\lvert \{ M_{\bf \tau }= N c_{\btau} \} \ket = \lvert
+,+,\ldots , + \ket$ belongs to ${\cal V}^s$ since its angular
momentum component along the $z$ axis equals $N/2$ and is
maximal. Moreover, the lowering operators $J^- _{\btau}$ leave
${\cal V}^s$ invariant and $ \lvert \{ N c_{\btau} - 2 d_{\btau} \}
\ket = \otimes (J^- _{\btau} )^{d_{\btau}} \lvert \{ N c_{\btau} \}
\ket$.  In addition, the equilibrium Gibbs vector belongs to ${\cal
V}^s$.  We shall see in the following that not only the Gibbs mode but
also all the eigenvectors corresponding to the smallest eigenvalues
lie in ${\cal V}^s$.

At infinite temperature, the system is purely diffusive and the
equilibrium state $\lvert 0 \ket$ equals the normalized sum of all
$\lvert {\bf S} \ket$s. The average values of the angular momentum read
$\bra 0 \lvert J^y \lvert 0 \ket = \bra 0 \lvert J^z \lvert 0 \ket =
0$ , $\bra 0 \lvert J^x \lvert 0 \ket = N/2$. For low-lying
excitations $\lvert \omega \ket$, that is plane waves on the
$N$-dimensional hypercube, the mean value of the angular momentum
$\bra \omega \lvert {\bf J} \lvert \omega \ket$ equals the one in the
equilibrium state up to non extensive terms. The same picture 
holds at finite temperature, but not at phase transition boundaries.  More
precisely, equilibrium or quasi-stationary modes $\lvert 0 \ket$ are
expected to correspond to well-defined direction of the angular
momenta ${\bf J}_{\btau}$ in the three-dimensional space. This direction
smoothly varies when exploring the excited modes slightly above $\lvert 0 \ket$
\footnote{Activated processes between wells can not be obtained this
way but they correspond to infinite time scales we do not investigate
here.}. To capture the spectrum of low-lying states, we therefore
proceed in two steps~: first identify the preferred directions of the
angular momentum corresponding to quasi-equilibrium modes and then
compute the fluctuations around these positions.

This approach, reminiscent of semi-classical approximations and spin
wave theory \cite{spinwaves}, may be made more precise in the
following way.  Consider the Holstein-Primakoff (HP) representation of the
angular momentum vectorial operator \cite{holpri} in terms of Bose
operators
\begin{eqnarray}
\label{holpri} 
L^{x}_{\btau} & = & \frac 12 N c_{\btau}  - a^{\dag}_{\btau} 
a_{\btau} \nonumber \\
L^{y}_{\btau} & = & \frac{1}{2i} 
\left[a^{\dag} _{\btau}\; \sqrt{
N c_{\btau}-a^{\dag}_{\btau} a_{\btau}}- \sqrt{
N c_{\btau}-a^{\dag}_{\btau} a_{\btau}} \;
a_{\btau}\right]\nonumber \\
L^{z}_{\btau} & = & \frac{1}{2}  
\left[a^{\dag} _{\btau} \; \sqrt{N c_{\btau}-
a^{\dag}_{\btau} a_{\btau}}+ \sqrt{
N c_{\btau}-a^{\dag}_{\btau} a_{\btau}} \;
a_{\btau}\right] \qquad ,
\end{eqnarray}
which is valid for fixed angular momentum $\frac 12 N c_{\btau}$ and
as long as the number of quanta does not exceed $N c_{\btau}$. The
Bose operators fulfill the usual commutation relation 
$[a_{\btau},a^{\dag}_{\btau'}]=\delta _{\btau,\btau'}$.
The large $N$ expansion will be consistent provided that we enforce 
the number of quanta to keep finite. To do so, we align the HP
representations (\ref{holpri}) along the quasi-equilibrium directions
of the angular momenta  
\begin{equation}
J_{\btau} ^x = \cos \theta _{\btau} \; L_{\btau} ^x -
\sin \theta _{\btau} \; L_{\btau} ^z \quad , \quad
J_{\btau} ^y = L_{\btau} ^y \quad , \quad 
J_{\btau} ^z = \sin \theta _{\btau} \; L_{\btau} ^x +
\cos \theta _{\btau} \; L_{\btau} ^z
\end{equation}
in every blocks. We may now expand the evolution operator $W^s= N
. W_0^s + \sqrt{N}. W_1 ^s + W_2 ^s + O(1/\sqrt {N})$ and find
\begin{eqnarray}
W_0^s &=& \frac 12 \sum _{\btau} c_{\btau} \left( \frac{\cos \theta
_{\btau} }{ \cosh K_{\btau}} + \sin \theta _{\btau} \tanh K_{\btau} -1
\right) \nonumber \\
W_1 ^s &=&  \frac{\sqrt{ c_{\btau}}}{2 \cosh K_{\btau}} (\cos
\theta_{\btau} \sinh K_{\btau} - \sin \theta _{\btau} ) \times 
\nonumber \\
& & \left( a^{\dag}_{\btau} +  a_{\btau} - 
\sum _{\mbox{\boldmath $\eta$}} \sum _{\mu =1 }^p 
\tau ^\mu \eta ^\mu \; 
\sqrt {c_{\mbox{\boldmath $\eta$}} c_{\btau} } \frac{\cos 
\theta _{\mbox{\boldmath $\eta$}}}{\cosh K_{\btau}} 
(a^{\dag}_{\mbox{\boldmath $\eta$}} +  a_{\mbox{\boldmath $\eta$}}) \right)
\end{eqnarray}
where
\begin{equation}
\label{defK}
K_{\btau} = \beta \sum _{\mu =1 }^p \tau ^\mu 
\sum _{\mbox{\boldmath $\eta$}} 
c_{\mbox{\boldmath $\eta$}} \; \sin \theta _{\mbox{\boldmath $\eta$}}
\; \eta ^\mu 
\end{equation}
Both $W_0^s$ and $W_1^s$ vanish when the angles are such as
$\sin \theta _{\btau} = \tanh K_{\btau}$ for all blocks
$\btau$. Expression (\ref{defK}) implies that the $2^p$ angles may be
computed through the knowledge of the $p$ parameters $\displaystyle{m
^\mu = \sum _{\mbox{\boldmath $\eta$}} c_{\mbox{\boldmath $\eta$}}
\sin \theta _{\mbox{\boldmath $\eta$}}\; \eta ^\mu}$. It turns out that
the resulting self-consistent equations for the $m^\mu$s are identical
to the equilibrium saddle-point conditions (\ref{mfeq}), as could be
expected from the above discussion.

As a consequence, the smallest eigenvalues gather together in
different families ${\cal F}$, in one to one correspondence with the 
stationary points of the free energy (\ref{freeenergy}). In each
family ${\cal F}$, we are left with the spin-wave part of the 
evolution operator
\begin{equation}
\label{w2}
W^s _2 = -\frac 14
\sum_{{\btau},{\btau}'} \left[ (a_{\btau}+a^{\dag}_{\btau })
( O^{2} - 1) _{{\btau},{\btau}'} 
(a_{{\btau}'}+a^{\dag}_{{\btau }'}) \right] 
-\sum_{\btau}\left[ a^{\dag}_{\btau}a_{\btau}+\frac 12 ( 1 - 
O_{{\btau},{\btau}} ) \right]
\end{equation}
where the $2^{p}\times 2^{p}$ matrix $O$ reads
\begin{equation}
\label{ofor} 
O_{{\btau},{\btau}'}=1_{{\btau},{\btau}'}-
\beta \sum_{\mu =1}^p \left( \frac{ \sqrt{ c_{\btau} }}
{ \cosh  K_{\btau} } {\btau }^\mu \right) \left( \frac{ \sqrt{ c_{{\btau}'}}}
{ \cosh  K_{{\btau}'} } {{\btau}' }^\mu \right) \quad .
\end{equation}
At least $2^p -p$ eigenvalues of $O$ equal one. A straigthforward
calculation show that the difference between the traces of the
$n^{th}$ powers of $O$ and of the free-energy Hessian matrix, see
(\ref{freeenergy}, \ref{mfeq}),
\begin{equation}
H_{\mu , \nu} = \left. \frac{\partial ^2 f}{\partial m^{\mu}
\partial m^{\nu}} \right| _{\cal F}
\end{equation}
equals $2^p-p$ for all integers $n$. In the above expression, the Hessian
depends on the particular family ${\cal F}$, i.e. saddle-point under
consideration. Therefore, the {\it a priori} non unit
eigenvalues of $O$, hereafter called $\lambda _j$s, $j =1, \ldots ,p$
are precisely the eigenvalues of the second derivatives matrix of the static 
free-energy.

To achieve the calculation of the spectrum of relaxation times, 
we now perform an orthogonal transformation over the Bose operators to
diagonalize $O$ and obtain
\begin{equation}
\label{Wdiag}
W^s _2= - \sum_{j=1}^{p} \left[  \frac 14 ( \lambda
_{j}^{2} - 1 ) (b_{j}+b^{\dag}_{j})^{2}
+  b^{\dag}_{j}b_{j}+ \frac 12 (1 -  \lambda
_{j}) \right]-\sum_{j=p+1}^{2^{p}} b^{\dag}_{j}b_{j} \quad .
\end{equation} 
The first term on the r.h.s. of (\ref{Wdiag})
includes $p$ interacting bosons which can be decoupled by means of 
a Bogoliubov transformation
\begin{equation}
\label{bogo}
\left( \begin{array}{c} c_j \\ c^{\dag}_j \end{array} \right)
= \left( \begin{array}{cc} \cosh \phi _j  & \sinh \phi _j  \\
\sinh \phi _j & \cosh \phi _j  \end{array} \right) \
\left( \begin{array}{c} b_j \\ b^{\dag}_j \end{array} \right)
\quad , 
\end{equation}
where $\phi _j = \frac 12 \ln \lvert  \lambda _j \lvert $.
We can finally rewrite $W^s _2$ in a completely diagonal form, 
\begin{equation}
\label{wfin}
W^s _2 =- \sum_{i=1}^{p} |\lambda _{j}| \big( c^{\dag}_{j}  c_{j} + \Delta
_{j} \big) - \sum_{j=p+1}^{2^{p}} b^{\dag}_{j}b_{j} 
\end{equation}
where $\Delta_{j}=1$ if $\lambda _{j} < 0 $ and $\Delta_{j}=0$ if
$\lambda _{j}> 0 $. In this form, the slow eigenmodes of the Glauber
matrix are seen to correspond to the eigenvectors of a quantum
harmonic oscillator and can be easily calculated \cite{dinamicaising}.
Note that transformation (\ref{bogo}) requires that no
eigenvalue vanishes. Indeed, for a marginal saddle-point ${\cal F}$, 
the large $N$ expansion has to take into account that the number
of quanta $a^{\dag} a$ becomes of the order of $\sqrt N $.

We can follow the same procedure to diagonalize $W^s$ in the
subspace labelled by angular momenta $\{J_{\btau} = \frac{1}{2} N
c_{\btau} - j_{\btau} \}$ where all $j_{\btau}$s are finite; we obtain
for $W^s _{\btau}$ expression (\ref{wfin}) plus the constant term $-
j_{\btau}$.  Therefore, the smallest eigenvalues do not belong to such
subspaces.  If the $j_{\btau}$s are not finite as $N$ goes to
infinity, the diverging evolution operators $W^s _0$ and $W^s _1$ can
never be simultaneously cancelled through adequate rotations.

Starting from a given initial condition, the evolution in the
parameter space $\{ m ^\mu \}$ will be deterministic \cite{vankampen}
and follow the macroscopic law
\begin{equation}
\label{dinmu}
\frac{d}{dt} m^{\mu }(t) = -\frac{\partial f }{\partial m^{\mu }} (\{
m ^\nu (t) \} ) \qquad ,
\end{equation}
giving rise to a gradient descent of the free-energy landscape down to
some local minimum. When the number $p$ of patterns increases, this
free-energy landscape becomes very complex, e.g. the number of
stationary points scales exponentially with $p$ \cite{amit}. Each
fixed point has a basin of attraction; a point for a local maximum, a
line for points with only one stable direction, \dots , a manifold of
dimension $N-I$ for a stationary point of index $I$. However, because
of pure entropic effects, some stationary points have much wider
basins of attractions that other ones, that is the number of
microscopic configurations associated to them may be extremely large.
For instance, a random and unbiased initial condition would fall into
the basin of the paramagnetic saddle-point $m ^\mu =0$, $\forall \mu$.

Generally speaking, when the starting point lies in a basin of
attraction of a stationary point $\{ m ^\mu \} _{\cal F}$ of the
free-energy, the system relaxes toward this stationary point. Later
on, the order parameters keep stucked to their saddle-point values for
all but finite times, even if ${\cal F}$ is unstable from a
thermodynamical standpoint. This description is reminiscent of the
{\it scenario} proposed in \cite{kurchan}.  The time scales we have
calculated here govern the $1/\sqrt N$ fluctuations around the
stationary point.  Fluctuations in the stable directions, that are
related to positive $\lambda _j$s, decay with relaxation times equal
to $1/\lambda _j$ and eventually equilibrate ($\Delta _j =0$).  As for
fluctuations along the unstable directions associated to negative
$\lambda _j$s, they exponentially grow $(\Delta _j=1$) with some
characteristic `escape' times $1/|\lambda _j |$. In both cases, the
number of affected spins is of the order of $\sqrt N$ and the
intensive quantities remain unchanged.

\end{document}